\documentclass[final,12pt]{elsarticle}
\usepackage[utf8]{inputenc}
\usepackage[english]{babel}
\usepackage{listings}
\usepackage{tabularx, caption}
\usepackage{graphicx,url}
\usepackage{multirow}
\usepackage{color}
\usepackage{enumitem}
\usepackage{booktabs}

\usepackage{graphicx}
\usepackage{amssymb}

\usepackage{lineno}

\journal{Journal}

\begin{document}

\begin{frontmatter}


\title{A Survey of Software Code Review Practices in Brazil}



\author[l1,l2]{Marcos Dósea}
\author[l1]{Claudio Sant'Anna}
\author[l2]{Ythanna Oliveira}
\author[l2]{Methanias Colaço Junior}

\address[l1]{Department of Computer Science, Federal University of Bahia, Salvador, Bahia, Brazil}

\address[l2]{Department of Information Systems, Federal University of Sergipe, Itabaiana, Sergipe, Brazil}

\let\thefootnote\relax\footnotetext{TR-PGCOMP-001/2020. Technical Report. Computer Science Graduate Program. Federal University of Bahia.}

\begin{abstract}
\emph{Context:} Software code review aims to early find code anomalies and to perform code improvements when they are less expensive. However, issues and challenges faced by developers who do not apply code review practices regularly are unclear. \emph{Goal:} Investigate difficulties developers face to apply code review practices without limiting the target audience to developers who already use this practice regularly. \emph{Method:} We conducted a web-based survey with 350 Brazilian practitioners engaged on the software development industry. \emph{Results:} Code review practices are widespread among Brazilian practitioners who recognize its importance. However, there is no routine for applying these practices. In addition, they report difficulties to fit static analysis tools in the software development process. One possible reason recognized by practitioners is that most of these tools use a single metric threshold, which might be not adequate to evaluate all system classes. \emph{Conclusion:} Improving guidelines to fit code review practices into the software development process could help to make them widely used. Additionally, future studies should investigate whether multiple metric thresholds that take source code context into account reduce static analysis tool false alarms. Finally, these tools should allow their use in distinct phases of the software development process.
\end{abstract}


\hyphenation{sys-te-ma-tic}
\hyphenation{de-ve-lopment}
\hyphenation{post-poning}
\hyphenation{stu-dies}
\hyphenation{a-reas}
\hyphenation{adjust-ments}
\hyphenation{par-ti-ci-pants}

\end{frontmatter}


\section{Introduction}\label{sec:introduction}

Software code review is a well-established practice used by leading companies and outstanding projects in the software industry \cite{Sadowski:2015:ICSE, McIntosh:2016:ESE, MacLeod:2018:IEEE}. It is recognized a cost-effective defect detection technique due to the early detection of bugs, when it is less expensive to fix \cite{Huizinga:2007:book}. The main objectives of code review are finding code anomalies (code smells), and performing code improvements in terms of readability, commenting, consistency, and dead code removal \cite{Bacchelli:2013:ICSE}.

Initially, formal inspections were proposed to perform code reviews \cite{Fagan:1976:ISJ}. Nowadays, modern code review (MCR) practices are more lightweight, informal, asynchronous, and supported by specialized tools. Many open-source software projects and companies such as Microsoft, Google, Facebook use MCR practices prior to merging new code into the main project codebase \cite{Rigby:2013:FSE, Balachandran:2013:ICSE}. Although manual code review is usually the main approach adopted to ensure source code quality, it is both error-prone and labor-intensive \cite{He:SERE:2013}. Therefore, efforts have been devoted to automating MCR practices.

Automated static analysis tools (ASATs) (e.g. PMD\footnote{http://pmd.sourceforge.net}, Checkstyle\footnote{http://checkstyle.sourceforge.net/}, SonarQube\footnote{https://www.sonarqube.org/} and NDepend\footnote{http://www.ndepend.com/}) are among the most popular MCR tools in industry to scan predefined problems and perform source code improvements in software systems \cite{Bessey:2010:JCA, Ayewah:2010:ISSTA, Zheng:2006:TSE}. Next to testing and manual code review, ASATs have become an important pillar of modern software quality assurance approaches \cite{Beller:2016:SANER}. However, some studies suggest that only very few software projects adopt these tools because programmers seem to not fully benefit from them \cite{Kumar:2013:FSE, Johnson:2013:ICSE, Beller:2016:SANER}.

An initial hypothesis for the low use of ASAT could be the lack of importance development teams and their companies assign to code review practices. Benefits of code review for software quality are already long recognized \cite{Fagan:1976:ISJ}. Surveys conducted with developers who regularly use code review practices indicate that developers spend 10-15 percent of their time in code reviews. Finding defects, performing code maintainability improvements, and knowledge transfer are the main reasons to perform code review \cite{Bacchelli:2013:ICSE, Bosu:2013:ESEM, Bosu:2017:TSE}. Additionally, other studies recognize that code anomalies impact the effort of different activities, such as editing, navigating, and reading of source code \cite{Soh:2016:SANER}. Although empirical evidence on the benefits of code reviews practices encourages their use, many software companies do not apply them regularly \cite{Johnson:2013:ICSE}. Previous surveys addressing difficulties to apply code review practices limited their target audience to software developers that regularly use these practices\cite{Kononenko:2016:ICSE, MacLeod:2018:IEEE, Christakis:2016:ASE}. Therefore, the issues and challenges faced by developers who do not apply these practices regularly are unclear.

Another possible concern faced by developers is to fit code review practices in their particular development process. Many studies about code reviews practices and tools focus on their correctness, completeness or performance in companies that already use these practices \cite{Christakis:2016:ASE, Bosu:2017:TSE}. But when an organization needs to integrate these practices in their development process other considerations need to be taken into account.  For example, Ayewah et al. \cite{Ayewah:2008:JIE} conducted some interviews to get qualitative feedback about systematic polices used by FindBugs users. They report that 76\% from users do not have systematic policies for using FindBugs and 81\% do not have a policy on how soon each FindBugs issue must be human-reviewed. Therefore, when code review practices are unclear into the development process they could be easily disregarded \cite{Lavallee:2015:ICSE}. The issues and challenges faced by development teams to integrate software code review practices into the software development process are unclear. A deeper understanding of these problems could guide researchers’ future work to offer alternatives and improve guidelines.

Another common challenge is to identify the best moments to apply code review with the support of ASATs \cite{Vassallo:2018:SANER}. Studies show that code anomalies may affect artifacts since their creation \cite{Tufano:2015:ICSE} and even the most experienced developers also introduce code anomalies \cite{Lavallee:2015:ICSE}. ASATs only launched by an integration server or by developers anytime they want may end up postponing important repairs or leading to lack of motivation and time to perform code reviews \cite{Sadowski:2015:ICSE}. It is valuable to fix potential code anomalies when they are introduced into source code because the context necessary to understand the anomaly is already in the developers’ working memory \cite{Johnson:2013:ICSE}. Recent studies propose to show immediate feedback to developers, providing the most relevant results quickly, and computing less relevant results incrementally later \cite{Do:2017:ISSTA}. Despite the perceived benefits from some initial evaluations \cite{Tymchuk:2018:ICPC}, a large-scale perception of practitioners about the most appropriate moments to check code anomalies is important to guide future research and the development of tools.

Finally, some studies have reported that current ASATs are prone to false-positive alarms. These false-positives correspond to warnings that manual inspection reveals no effect on the software quality and on maintenance effort \cite{Olbrich:2010:ICSM, Khomh:2011:JSS, Sjoberg:2013:TSE, Yamashita:2013:ICSM, Palomba:2014:ICSME, Hozano:2018:IST}. Despite developers being able to eliminate many defects using the warnings produced by these tools, the overload of warnings and the way in which they are presented are pointed out as the main barriers to the consistent and widespread use of ASATs \cite{Ayewah:2010:ISSTA, Johnson:2013:ICSE}. Some studies indicate that only 6\% to 22\% of warnings are removed in the context of code reviews \cite{Kim:2007:FSE, Panichella:2015:SANER}. Analyzing warnings is a time-consuming activity. For instance, a study conducted at Google indicated that, on average, eight minutes are required to manually triage each static analysis warning \cite{Ruthruff:2008:ICSE}. One line of research to reduce these false alarms is improving the accuracy of metric thresholds used by popular metric-based ASATs \cite{Marinescu:2004:detectionstrategies, Arcoverde:2012:RSSE, Balachandran:2013:ICSE, Oizumi:2016:icse}. These tools usually use generic metric thresholds for classifying source code elements (such as classes and methods) of one or more systems into categories (e.g. low or high) \cite{Lanza:2006:oo-metrics, Alves:2010:ICSM, Ferreira:2012:JSS, Oliveira:2014:CSMR, Fontana:2015:WETSOM, Vale:2015:SBES}. For instance, Lanza and Marinescu \cite{Lanza:2006:oo-metrics} classify as \emph{long} any method that has more than 20 lines of code (LOC) in Java systems. In this case, 20 is used as a generic threshold for LOC.

However, some studies suggest the major reason for the occurrence of false positive and false negatives warnings is the lack of context for metric thresholds \cite{Zhang:2013:ICSM, Aniche:2016:SCAM, Sharma:2018:JSS, Dosea:2018:ICPC}. For example, for systems that follow the layered architectural style, there might be differences between the average source code complexity of classes belonging to the View layer and the Business layer. The business logic implementation is often more complex than View logic implementation.  A generic threshold value might be low for classes in a layer or high for classes in another layer. Too low or too high thresholds may lead to false code smell alarms (false positives) or may hide potential code smells (false negatives). Therefore, applying a single generic threshold to evaluate classes in these distinct layers may not make sense. An obvious solution would be to propose multiple threshold values for each metric. In our example, we could offer a distinct metric threshold for each architectural layer of the system. In this case, architectural layers would be used as context to define multiple metric thresholds. However, the practitioners' perception about multiple metric thresholds for source code evaluation is also unclear.


In this paper, we conducted a large-scale web-based survey with 350 Brazilian practitioners that most often do not have well-established practices to review source code. The main goal is to assess whether ongoing researches are addressing the same issues and challenges practitioners face to apply code review practices in their development process. Our results show that, although 84.85\% of respondents claimed to use at least one code review practice and 54.85\% declared to use tools that support code review, they do not apply these practices and tools regularly. The results also show that we can not justify this lack of regularity to the level of importance developers and companies give to code review practices. The respondents also pointed out that fitting these practices to the software development process and configuring the tools are key challenges they face. Another finding is that developers agreed that using multiples metric thresholds for each metric, configured according to the class design context, could decrease the number of false alarms. Finally, respondents claimed that the best time to warn developers about code anomalies is during the typing of source code.
    
The remainder of this paper is structured  as follows. Section \ref{sec:studySettings} presents details of the methodology used to perform the survey, including purpose, instrument, and data collection. Section \ref{sec:results} discusses results of the survey, including background information of the respondents and issues and problems reported by practitioners to apply code review practices. Section \ref{sec:threats} discusses threats to validity. Section \ref{sec:relatedWorks} brings the related works. Finally, section \ref{sec:conclusion} brings the conclusion, remarks, and future work.
\section{Study Settings}\label{sec:studySettings}

In this section we present the goal and research questions of our survey (Section \ref{sec:survey-goal}). We then discuss the statistics on the sample and the design of the survey questionnaire in Section \ref{sec:survey-design}. Section \ref{sec:survey-execution} discusses the survey execution and data collection procedures. In Section \ref{sec:survey-analysis} we discuss the analysis methodology used to answer the proposed research questions.

\subsection{Goal and Research Questions}\label{sec:survey-goal}

The goal of this survey is to characterize code review practices that software developers in Brazil use to ensure source code quality. Our purpose is identifying trends, difficulties and challenges to use these practices which could be addressed by researchers both in Brazil and worldwide. Based on this goal, we raise the following research questions (RQ).   
\begin{itemize}
	\item \emph{RQ1: What are the code review practices adopted by developers in Brazil to evaluate source code quality?} With this research question we aim to gain a broader view of the most used code review practices. We also intend to assess how often these practices are used.
	\item \emph{RQ2:  How important do developers in Brazil perceive code review practices?} With this research question, we aim to understand which level of importance developers in Brazil assign to code review practices, since this perception could affect the adoption of such practices.
	\item \emph{RQ3: What difficulties do developers in Brazil face to use automated static analysis tools to support code review?} Several studies discussed the completeness of ASATs to point out relevant alarms, but only few discuss issues and challenge to fit them in the software development process \cite{Vassallo:2018:SANER}. In this research question, our goal is identifying problems that may make developers in Brazil avoid using automated static analysis tools to support code review.
	\item \emph{RQ4: What is the developers' perception about evaluating source code using multiple threshold values for each metric?} Most ASATs rely on metrics and thresholds to point out pieces of low-quality source code. For a metric, ASATs usually use a single generic metric threshold to evaluate all system classes. However, calibrating metric thresholds without taking context into account can increase the number of false-positive warnings \cite{Sharma:2018:JSS}. In fact, recent studies claim that considering context factors to define multiple thresholds could improve accuracy and reduce false-positive alarms \cite{Zhang:2013:ICSM, Aniche:2016:SCAM, Dosea:2018:ICPC}. In this research question, we aim to evaluate developers' perception about single and multiple metric thresholds. To make the notion of context more concrete to the respondents, in our questionnaire we suggested two context factors that could be used to define multiple metric thresholds: (i) the architectural layer of a class, and (ii) the main business entity handled by a class. The hypothesis about business entities is that some of them may be simpler to be handled than others. For example, considering a library management system, the business entity \emph{author of books} is usually simpler to be handled by the system than \emph{books} and \emph{borrowing of books} entities that usually involve more data and more complex business operations. We suggested these two factors based on factors considered in previous studies \cite{Aniche:2016:SCAM, Dosea:2018:ICPC}, but they may not be the only factors that affect metric thresholds.
	
	\item \emph{RQ5: What is the best time in the software development process to conduct automatic code review practices?} This research question investigates what time of the software development process developers in Brazil consider as the most appropriate moment to use automated static analysis tools. Although we know that code review should be conducted in various moments, we intend to evaluate whether the current tools provide resources to be executed according to the preference of the practitioners.
\end{itemize}

We used each RQ to derive one or more “survey questions” detailed in the following section. The exploratory nature of the survey questions aimed to clarify the current issues faced by practitioners to perform source code analysis. Additionally, they enabled us to seek new insights and generating hypotheses for future studies.

\subsection{Survey Design}\label{sec:survey-design}

We used guidelines reported by \cite{Kitchenham:2002:survey, Kasunic:survey:2005, Linaker:survey:2015} to design the survey. Surveys have been used in empirical software engineering investigations to learn about the state of the practice, identify improvement potentials, or investigate the acceptance of a technology insight \cite{Punter:2003:ISESE}. We describe below the sampling method, survey questions designed, execution process, and analysis methodology.
 
\subsubsection{Population and Sampling Method}

In our study, the target population is formed by Brazilian software practitioners engaged on the software development industry. We did not find official data in Brazil about our target population. For this reason, we use a non-probabilistic sampling, used by researchers when systematic probabilistic sampling is not possible. However, a non-official estimate published by SOFTEX (Association to Promote the Excellence of Brazilian Software)\footnote{https://www.softex.br/} and widely used by Brazilian press estimates 570 thousand IT (Information Technology) Brazilian professionals in 2018, working in a wide range of areas, including telecommunications, networks and software development. A total of 411 respondents started the questionnaire, whereof 350 (85,15\%) completed all mandatory questions. We then used the R tool to determine the level of confidence and the margin of error of the sample considering this total estimate of IT professionals. We obtained a confidence level of 95\% and a margin of error of 5.24\%. This means that, if we undertake 100 surveys for the same purpose and with the same methodology, in 95 of them the results would be within the margin of error. The size of our sample is considerable when compared with previous surveys in software engineering, especially Brazilian surveys \cite{Agner:2013:JSS}, which reach much smaller numbers of respondents and, therefore, much larger margin of error.

We decided to use a self-recruited survey, in which the respondents get to know somehow about the survey and decide to participate. The main advantage of a self-recruited survey is that respondents are attracted already by the topic of the survey \cite{Punter:2003:ISESE}. We use an online questionnaire created by means of the SurveyMonkey tool\footnote{http://www.surveymonkey.com} to collect the data. The motivation for using an online questionnaire was to maximize coverage and participation. It also allows an easier data entry from the respondent perspective, a simpler data collection from the researcher perspective and is less error-prone \cite{Punter:2003:ISESE}.


\subsubsection{Survey Questions}\label{sec:survey-questions}

We designed our survey questionnaire aiming to answer the research questions (Section \ref{sec:survey-goal}) and to characterize the respondents. We were also concerned to avoid a large questionnaire, which would take a long time from the respondents. Studies showed that short questionnaires have a higher response rate in comparison to long ones \cite{Punter:2003:ISESE, Smith:2013:Chase}. Thus, we consider five minutes as target time for the participants to answer all questions.

To achieve this goal, we elaborated clear and objective questions. Additionally, we conducted two pilot studies before coming up with the final version of the questionnaire. The first one involved five undergraduate students attending the last period of an information systems course. The second one involved 15 experienced professionals (5 to 20 years working in software development industry) who attended a course given by one of the authors. The pilot studies involved these two group of developers because we also aimed to assess whether the proposed questionnaire was suitable for both new developers and expert developers. During the pilot studies, we monitored time, questions the participants made, and register misunderstandings due to question formulation. Everyone was able to answer the questionnaire within 4 to 5 minutes. Then we discussed with them their understanding about each question. As a result of the pilot studies, we made some adjustments in the vocabulary used in the questions. Some participants raised doubts about practices or tools mentioned in some questions. Thus, we added examples to illustrate them and make the questions clearer.

The questionnaire is organized into four sections. Each section was showed to the respondent in a distinct Web page. All the pages had the same title that reflected our main research goal: ``Which practices do you use to evaluate the quality of source code?''. The first section informs to the respondents that the survey would required around five minutes from them. It also that informs that the survey has eleven questions. In addition, the first section gives a brief explanation about source code review, and source code review practices and tools.

Table \ref{table:questions} lists the survey questions. For each question, it informs the questionnaire section where the question appears, the related research question, the question itself (translated from Portuguese into English), the type of allowed answer and the number of respondents. Overall, the questionnaire consists of seven questions, plus four background questions. Some questions allow the respondent to select a single answer (S) and other questions allow the respondent to select multiple answers (M). Additionally, some questions allow an open answer (O) where the respondent can write a free text response to add an answer not included in the list of answer options. For instance, the first question is related to research question RQ1 and appears in the second section of the survey. It allows multiples answers and an open answer. And 411 respondents answered it. 

The second section of the survey has four questions related to the first two research questions. To answer RQ1 it includes two questions about the code review practices the respondent uses and with which frequency he or she uses them. To answer RQ2 the questionnaire contains two questions about the importance given by the respondent and the company to the use of code review practices. All 411 respondents who started the survey advanced through the first and second sections, answering all the questions. We elaborated the simplest questions in the second section aiming to encourage the progress to the other survey questions.  

The third section contains three questions aiming to identify issues and challenges faced by practitioners to use code review techniques. Each question is associated to one research question (RQ3, RQ4, and RQ5). The third section contains more reflective questions and obtained high dropout rate 56 out of 411 (13.62\%) of the respondents.

Finally, the four section is concerned with the background of the participants, including questions about experience with software development as well as about the number of developed systems that they already worked with. We structured the background section at the end aiming the respondent to focus from the early stages on the main objectives of the survey \cite{Seaman:1999:TSE}. We believe that this approach also influenced the high rate of responses we obtained. In the last section, there were only four dropouts. In summary, the median number of responses per question was 411 for RQ1 and RQ2 questions, 355 for RQ3, RQ4 and RQ5 and 350 for background questions.


\begin{table*}[t]
\caption{Questionnaire - Questions (S/M/O Stands for Single, Multiple, or Open answer).}
	\label{table:questions}
	\resizebox{\textwidth}{!}
{
\begin{tabular}{cclcc}
\hline
\multicolumn{1}{l}{Section} & \multicolumn{1}{l}{\begin{tabular}[c]{@{}l@{}}Research Question\\ or Background\end{tabular}} & Question & S/M/O & \#Respondents \\
\hline
2 & RQ1 &  \begin{tabular}[c]{@{}l@{}}What code review practices do developers in your company \\ use to analyze the quality of source code?\end{tabular} & M, O & 411 \\
2 & RQ1 & How often do you apply code review practices? & S & 411 \\
2 & RQ2 & What importance do you give to code review practices? & S & 411 \\
2 & RQ2 & What importance does your company give to code review practices? & S & 411 \\
3 & RQ3 & What difficulties do you have to use code review tools? & M, O & 355 \\
3 & RQ4 & \begin{tabular}[c]{@{}l@{}}Automated Static Analysis tools usually use a single metric threshold to evaluate all system \\ classes. Considering a three-tier system (GUI, Business and Persistence), what is your opinion \\ about the threshold values that should be used to evaluate classes in each of these three tiers?\end{tabular} & S & 355 \\
3 & RQ5 & \begin{tabular}[c]{@{}l@{}}What is the best moment to warn software developers about code anomalies? \end{tabular} & S, O & 355 \\
4 & Background & What is your highest academic degree? & S, O & 350 \\
4 & Background & What is your current role in the company? & S,O & 350 \\
4 & Background & How much experience do you have in software development? & S & 350 \\
4 & Background & How many systems have you developed or performed maintenance tasks? & S & 350
\\
\hline
\end{tabular}
}
\end{table*}

\subsection{Execution}\label{sec:survey-execution}

We made the survey available through an online questionnaire created by means of the SurveyMonkey\footnote{http://www.surveymonkey.com} tool. The sampling considered was 350 respondents. Given the lack of any  reliable data about the population of Brazilian software developers, we selected developers to participate in the survey as follows:

\begin{itemize}
\item We sent invitations through Google, Linkedin and Facebook software developers online forums and private groups, such as, C\# Brazil \footnote{https://www.facebook.com/groups/csharpbrasil}, Java Brazil\footnote{https://www.facebook.com/groups/JavaBr/}, Web Development Brazil\footnote{https://www.facebook.com/groups/desenvolvimentoweb/} and Android Brazil\footnote{https://www.facebook.com/groups/androidbrasiloficial/}. We were also invited to publish the survey on the main page of an important Brazilian software quality website\footnote{http://qualidadedesoftware.com.br/}. The publication on the site was available for approximately one month. 
\item We sent invitation emails to the main researchers associated with this field of study in Brazil. We requested that they submit the survey to students and professionals engaged in the area of software development.
\end{itemize}

The survey was conducted from January 2015 until September 2018. Most of the questionnaires (363 out of 411) were responded by accessing the online survey in January 2015, which was the most intense period of dissemination of the survey. Aiming to improve the sample reliability, we performed a new disclosure in August 2018, totaling 411 forms. We noticed that the new responses did not significantly alter the results obtained with the first responses, demonstrating that the sample would already be large enough to obtain reliable results.

\subsection{Analysis methodology}\label{sec:survey-analysis}

We address the research questions discussed in Section
\ref{sec:survey-goal} through descriptive statistics and using statistical hypothesis testing to conduct a cross-factor analysis of source code review practices and practitioners' background. We aim to understand the challenges of using source code review practices associated to each class of practitioners. 

We apply the chi-square independence test to conduct analysis in research questions working with categorical variables. This test evaluates if two categorical variables are related in some population. We highlight that a chi-square test assumes that observations are independent of one another and that each observation can be assigned to one and only one category. For example, we used the hypothesis test to evaluate whether the level of practitioners experiences influenced the definition of the best time to evaluate the quality of the source code. For the effect size of a chi-square independence test, we use the Cramér’s V (a sort of Pearson correlation for categorical variables) because we have at least one nominal variable in all performed tests. Cramér’s V takes on values between 0 and 1. Values closer to 0 indicate a weak association and values closer to 1 indicate a strong association among variables.

In order to verify whether the differences of developers perception about the importance of code review practices between them and companies are statistically significant, we use the non-parametric Wilcoxon Rank Sum test [142] with 5\% significance level (i.e. p-value $<$ 0.05). We also estimate the magnitude of the observed differences using Cliff's $\delta$ \cite{Cliff:1993:PB}, a non-parametric effect size measure for ordinal data. We follow the guidelines of Romano \emph{et al.} \cite{Romano:2006:EffectSize} to interpret the effect size based on Cliff's $\delta$. Considering $\delta$ as effect size, ranging from -1 to 1, $\mid\delta\mid < $0.147 means negligible effect, $\mid\delta\mid < $0.33 means small effect, $\mid\delta\mid < $0.474 means medium effect, and $\mid\delta\mid >= $0.474 means large effect.  Effect sizes must be judged according to the context and even small effects might be of practical importance \cite{kampenes:2007:effectsize}.
\section{Results and Discussion}\label{sec:results}

From 411 professionals who accessed the Web questionnaire, 350 filled the entire questionnaire, yielding a 85.15\% response rate. This rate is higher than other on-line surveys in software engineering \cite{Punter:2003:ISESE}. We considered that the following reasons were determinant for this high rate: (i) the short time required to answer the survey (5 minutes), emphasized in the first screen, (ii) the use of simple and objective language to invite the respondents, and (iii) the use of examples to explain some terms the respondents might not be familiar with. Due to the type of sampling, we are not able to determine the number of respondents who received the questionnaires, therefore, the response rate takes into account the respondents who really opened the questionnaire.

This section initially shows the background of the respondents, and it presents and discusses results regarding the five research questions. 

\subsection{Respondent background}

Regarding the predominant role, Table \ref{tab:function} shows that 96 (27.43\%) respondents are programmers and 97 (27.71\%) are software engineers. Thirty-three (9.43\%) respondents are software architects. Fifty-four (15.43\%) declared themselves as quality analysts, role which usually has source code quality analysis as one of its tasks. Programmers, software engineers, software architects and quality analysts are roles concerned somehow with source code maintenance. Finally, 70 (20.00\%) respondents indicated other roles, for example, technical leader of programmers. We observed that 28 out of 70 respondents that indicated other roles also performed tasks related to source code maintenance. In summary, 308 (88\%) respondents play a role associated with source code maintenance and thus, should concern with source code quality. Other roles cited by respondents are related to software testing, project management and company direction.

\begin{table}[h]
\centering
\caption{Predominant role of the respondents.}
\label{tab:function}
\begin{tabular}{@{}lrr@{}}
\toprule
\textbf{Role} & \textbf{Respondents} & 
\textbf{(\%)} \\ \midrule
Programmer & 96 & 27.43\% \\
Software engineer & 97 & 27.71\% \\
Software architect & 33 & 9.43\% \\
Quality analyst & 54 & 15.43\% \\
Other (please specify) & 70 & 20.00\% \\ 
\bottomrule
\end{tabular}
\end{table}

In order to understand the respondents’ educational background, we asked them to inform their highest academic degree. Table \ref{tab:educational} shows that 175 out of 350 (50\%) respondents are bachelor and 105 (30\%) has professional certificate. We also observed 34 (9.71\%) respondents have master degree and seven have doctoral degree (2\%). Only 29 (8.29\%) respondents claim to have associate degree. In Brazil, associate degree varies between 2 to 3 years of full-time studies. This degree provides highly specialized knowledge (e.g. Web developer). Regarding professional certificate, it requires a previous bachelor degree for admission and performs a specialization course in one area of study, mostly addressed to professional practice. The results indicate a high educational level of the respondents of which 91.71\% have at least bachelor degree.

\begin{table}[tbp]
\centering
\caption{Respondents' highest academic degrees.}
\label{tab:educational}
\begin{tabular}{@{}lrr@{}}
\toprule
\textbf{Academic Degree} & \textbf{Respondents} & 
\textbf{(\%)} \\ \midrule
Associate & 29 & 8.29\% \\
Bachelor & 175 & 50.00\% \\
Professional Certificate & 105 & 30.00\% \\
Master & 34 & 9.71\% \\
Doctoral & 7 & 2.00\% \\ \bottomrule
\end{tabular}
\end{table}

Regarding work experience, Table \ref{tab:experience} shows that 112 (32\%) respondents has 2 to 5 years of work experience. Others 87 (24.86\%) respondents claim to have 5 to 10 and 92 (26.29\%) more than ten years of work experience. Finally, 59 (16.86\%) respondents have less than two years of experience. That means the majority of respondents are practitioners who have a reasonable level of experience in software development tasks.

\begin{table}[tbp]
\centering
\caption{Work experience.}
\label{tab:experience}
\begin{tabular}{@{}lrr@{}}
\toprule
\textbf{Years}
& \textbf{Respondents}
& \textbf{(\%)} \\ \midrule
0 - 2 & 59 & 16.86\% \\
2 - 5 & 112 & 32.00\% \\
5 - 10 & 87 & 24.86\% \\
$>$ 10 & 92 & 26.29\% \\
\bottomrule
\end{tabular}
\end{table}

To be sure about the level of experience level with development tasks, we also asked about the number of systems the respondents work on. Table \ref{tab:systems} shows that 134 (38.51\%) respondents claimed to have contributed with more than ten software projects and 75 (21.55\%) between 6 to 10 software projects. Also, 75 (21.55\%) respondents claimed to have contributed with 3 to 5 software projects, and only 64 (18.39\%) respondents claimed to participate in less than three software projects. These results illustrate that most of the respondents have a reasonable experience of which 81.61\% contribute with at least three software projects.

\begin{table}[tbp]
\centering
\caption{Number of Developed Systems.}
\label{tab:systems}
\begin{tabular}{@{}lrr@{}}
\toprule
\textbf{Systems} 
& \textbf{Respondents}
& \textbf{(\%)} \\ \midrule
0 - 3 & 64 & 18.39\% \\
3 - 5 & 75 & 21.55\% \\
6 - 10 & 75 & 21.55\% \\
$>$ 10 & 134 & 38.51\% \\
\bottomrule
\end{tabular}
\end{table}

\begin{figure}
	\centering
\includegraphics[width=1.0\linewidth]{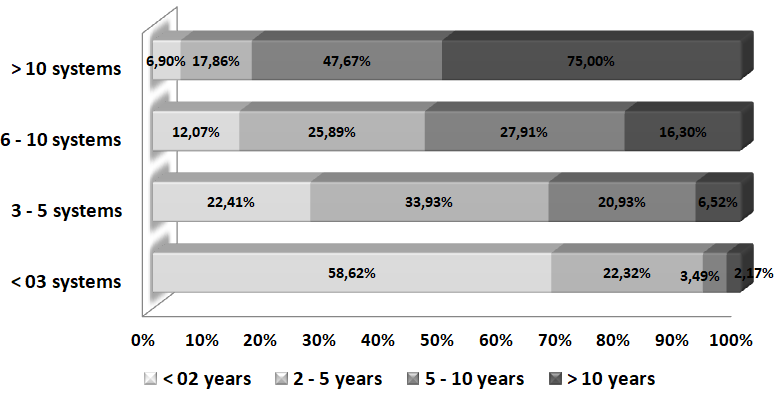}
	\caption{Number of developed systems according to the work experience of the respondents}
	\label{fig:systemsExperience}
\end{figure}

We conducted an analysis to verify to which degree practitioners' work experience is associate with the number of developed systems. This analysis is important because a strong association implies that we do not need to take both variables into account during our data analysis. To do that, we cross-analyzed the answers of the two questions. First, we applied the chi-square independence test, which is a statistical test used for evaluating if two categorical variables are related in some population. With this test we aimed at verifying whether the following null hypothesis is rejected: \emph{There is no association between number of systems developed and practitioners' working experience}. The test results ($\chi^2(9) = 160.49$, \emph{p}-value $< 0.001$) rejected this null hypothesis because the $\chi^2$ value with 9 degree of freedom is larger than the critical value at the 0.05 level ($16.92$). This means that there is an association between the number of developed systems and the practitioners' work experience. To measure the strength of this association (effect size) we used Cramér’s V. We obtained Cramér’s V = 0.153, which indicates a very weak association, since Cramér’s V takes on values between 0 and 1. Figure \ref{fig:systemsExperience} illustrates this association. We can see that there are some practitioners with little work experience who have already worked on a large number of software projects. In addition, there are practitioners with large work experience who have been involved in only few software projects. For this reason, the impact of these two variables on the other questions was evaluated separately.

We used the data obtained from the background questions to evaluate whether respondent's profile influenced the result about the research questions presented in the next subsections.

\subsection{RQ1: What are the code review practices adopted by developers in Brazil to evaluate source code quality?}

\begin{figure}
	\centering
	\includegraphics[width=1.0\linewidth]{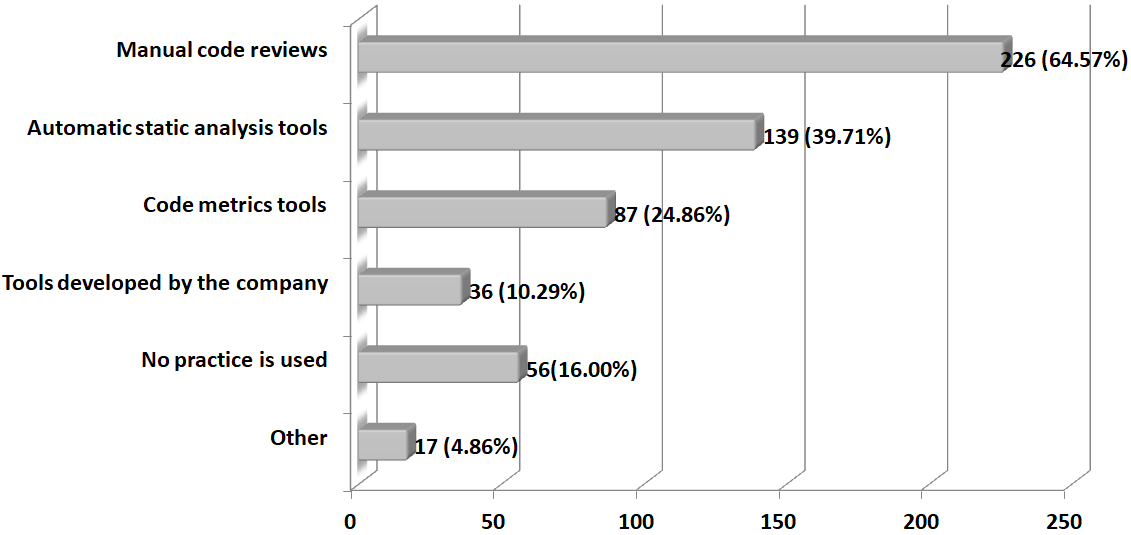}
	\caption{Code review practices adopted by the Brazilian companies to assess the quality of source code.}
	\label{fig:codeReviewPractices}
\end{figure}

To answer this research question, we analyze data from the first two survey questions (Table \ref{table:questions}). First, we asked the respondents about the practices they use to review source code quality in their companies. Respondents could select one or more answer options. Figure \ref{fig:codeReviewPractices} shows that 226 (64.57\%) respondents claim to use manual code reviews, and of those, we observed that 96 (27.42\% of 350) use manual code review exclusively. We observed that 139 (39.71\%) respondents declared to use automated static analysis tools (e.g. PMD, CheckStyle, ReSharper, FxCop), 87 (24.86\%) use code metric tools and 36 (10.29\%) use tools developed by their company itself. Finally, 56 (16.00\%) respondents declared that do not perform source code review and 17 (4.86\%) respondents claimed to use others practices. Some respondents indicated to use some specific code analysis tools (e.g. SonarQube and CodeClimate). Others also indicated the use of unit and integration test tools, which are usually not considered as source code review tools. We also identified 67 (19.14\%) respondents that claimed to use more than one tool to source code review (e.g. code metrics tool and tool developed by the company). Therefore, although manual code reviews are still the practice most often cited (64.57\% from respondents), we identified 192 (54.85\%) respondents that use at least one tool to perform code review. Finally, the results indicated that 297 (84.85\%) respondents use at least manual or automated practice to source code review, showing that respondents have reasonable level of knowledge about code review practices. 

\begin{figure}
	\centering
	\includegraphics[width=0.92\linewidth]{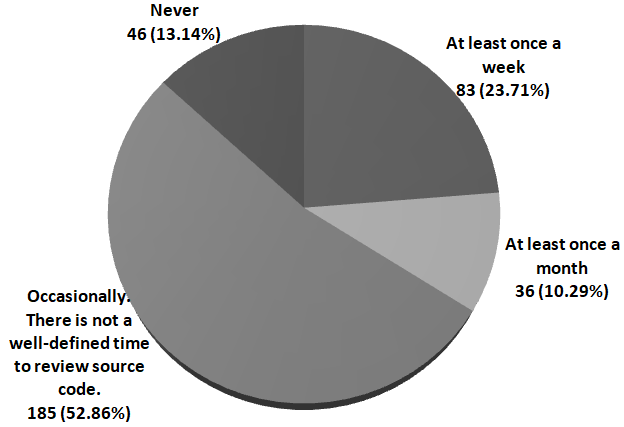}
	\caption{Frequency of use of each code review practice.}
	\label{fig:codeReviewFrequency}
\end{figure}

Secondly, we asked respondents how often they use code review practices. Our goal is to complement the results of the previous question by verifying how often these practices are actually applied in the development process. Figure \ref{fig:codeReviewFrequency} illustrates that 185 out of 350 (52.86\%) respondents declared that there is not a well-defined time to review source code, but occasionally revise it. We observed 36 (10.29\%) respondents that declared reviewing source code at least once a month and 46 (13.14\%) that claimed never reviewing the source code. Only 83 (23.71\%) respondents declared reviewing source code at least once a week. These results are not aligned the idea that software teams should meticulously review each change to source code to ensure quality standards \cite{Tanaka:1995:ICSE}. A survey conducted with Microsoft and OSS developers that adopt code review practices regularly found that they spend approximately six hours per week in code review \cite{Bosu:2013:ESEM}. Comparing these results with our survey results, we can say that 76.29\% of Brazilian practitioners do not perform code review regularly.

\begin{figure}
	\centering
	\includegraphics[width=1.0\linewidth]{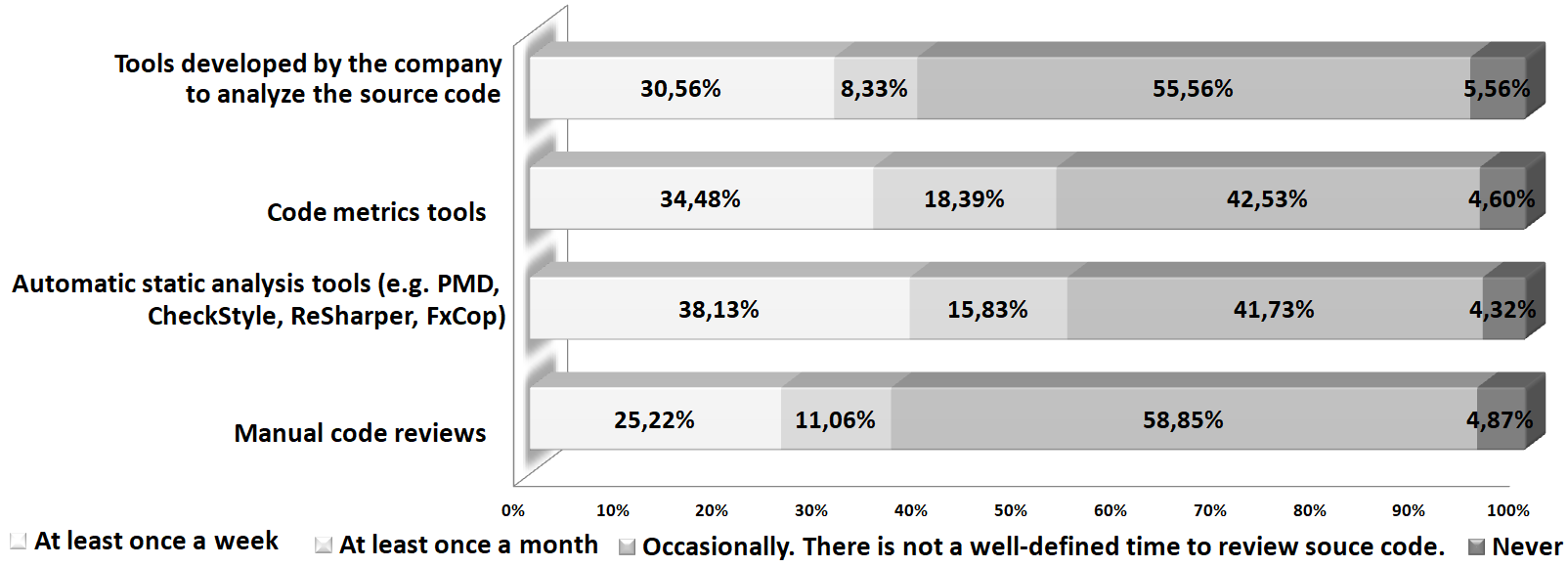}
	\caption{Frequency of use of the source code review practices.}
	\label{fig:practicesFrequency}
\end{figure}

Finally, we conducted a cross-analysis between answers about code review practices (first question) and answers about frequency of code review (second question). Figure \ref{fig:practicesFrequency} shows that, although 226 respondents (64.57\% of the total of respondents) claim to perform manual code review, only 57 out of them (25.22\% of 226) do it at least once a week, and only 25 (11.06\% of 226) do it at least once a month. In fact, most of those who claimed to perform manual code review, 133 respondents (56.85\% of 226), do not have a defined period to do that. In addition, 11 (4.87\%) respondents claimed to never review source code because they are not responsible for this task in their company. These results are similar to the other source code review practices. Regarding respondents who claim to use automated static analysis tools, 58 out of 139 (41.73\%) do not have a defined period for doing it and 6 (4.32\%) never review source code. Regarding respondents that who claim to use code metrics tools, 28 out of 69 (40.57\%) do not have a defined period to conduct reviews. Finally, regarding those who claim to use tools developed by their company, 21 out of 33 (63.63\%) also do not have a defined period to review code or never review code. Rigby et al. \cite{Rigby:2008:ICSE} show that early and frequent reviews of small, independent and complete contributions is an efficient and effective source code quality control practice in Apache projects. In this context, our findings suggest that the software development processes of several Brazilian companies needs improvement in terms of systematic adoption of code review in order to benefit of such practice.

In summary, based on the responses to the first two questions of our survey, we can answer RQ1 as follows:

\medskip

\fbox{\begin{minipage}{30em}
\textit{
Software code review practices are well disseminated among Brazilian practitioners. However, they are not applied regularly in the software development process. Thus, the positive impact of these practices on source code quality may not be perceived by development teams and companies.}
\end{minipage}} \linebreak

\subsection{RQ2: How important do developers in Brazil perceive code review practices?}

For code review practices to be applied regularly, recognizing its importance is crucial. In this sense, our second research question aims to figure out what level of importance Brazilian practitioners and their companies give to code review practices.

\begin{figure}
	\centering
	\includegraphics[width=0.75\linewidth]{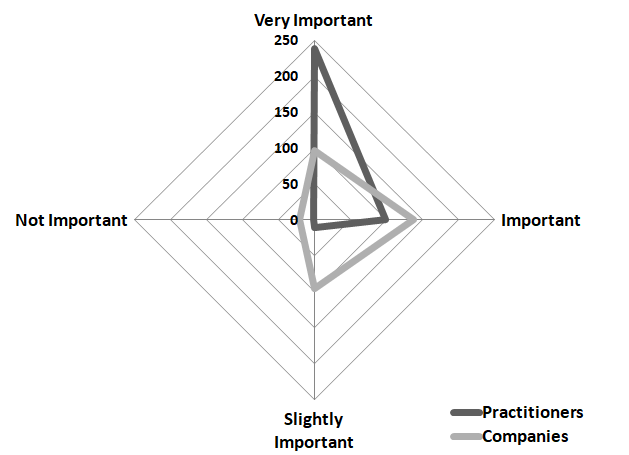}
	\caption{Importance of code review practices.}
	\label{fig:practicesImportance}
\end{figure}

Figure \ref{fig:practicesImportance} illustrates the level of importance both practitioners and companies give to code review practices. It puts together the answers for the third and forth questions of our survey, which are: \emph{What importance do you give to code review practices?} and \emph{What importance does your company give to code review practices?} We can notice that, for 239 (68.29\%) respondents, code review practices are very important and, for 98 (28\%), they are important. Only 11 (3.14\%) and 2 (0.57\%) respondents declared to consider code reviews as slightly important or as not important, respectively. Regarding the level of importance attributed by companies, 97 (27.71\%) and 137 (39.14\%) respondents declared to believe that their companies consider code review practices as very important or as important, respectively declared their companies give. On the other hand, 95 (27.14\%) respondents declared their companies give slightly importance. Finally, 21 (6.00\%) claimed to believe that their companies do not give any importance to code review practices. 

This data allows us to raise the hypothesis that developers believe that companies give less importance to code review practices than they do. Thus, we conducted a statistical analysis to test this hypothesis. We cross-analyzed the answers of the two questions. First, we converted the responses to the Likert scale, primarily used in questionnaires to assess subject’s perception. Each level of the scale corresponds to a numeric value \cite{Jamieson:2004:ME}. We assigned the following values: 4 - very important, 3 - important, 2 - slightly important, and 1 - not important. Then, we evaluated whether the data follows a normal distribution or not. We used the Shapiro-Wilk test because it is considered the most powerful normality test for sample sizes greater than 30 \cite{Kitchenham:ESE:2017}. The Shapiro-Wilk test indicated that the two set of responses about the level of importance given to code review practices do not follow a normal distribution (\emph{p}-value $<$ 0.001). Therefore, we used the Wilcoxon signed-rank test \cite{Corder:2014:book}, a nonparametric significance test to investigate if a variable tends to have statistical significant higher values than another one. As result, the test rejected the following null hypothesis: \emph{Practitioners perceive that companies give the same level of importance to code review practices as them}. In addition, we obtained Cliff's $\delta = 0.4843$, which corresponds to large effect size. This result means that, according to practitioners’ perception, their companies give less importance to code review practices than they do.

Finally, we investigate if the level of importance assigned by practitioners to code review practices are associated to their work experience and number of systems they developed. First, we cross-analyzed the answers applying the chi-square independence test aiming to verify whether the following null hypothesis is rejected: \emph{There is no association between the level of importance assigned by practitioners to code review practices and practitioners' work experience}. The test results ($\chi^2(9) = 14.27$, \emph{p}-value=$0.113$) did not reject this null hypothesis because $\chi^2$ value with 9 degree of freedom is lower than the critical value at the 0.05 level ($16.92$). This means that there is no association between the level of importance assigned to code review practices and practitioners' work experience. To strengthen this result we calculated the effect size and obtained Cramér’s V = 0.013, which indicates a very weak association. Second, we cross-analyzed the answers applying the chi-square independence test aiming to verify whether the following null hypothesis is rejected: \emph{There is no association between level of importance assigned by practitioners to code review practices and the number of systems they developed.}. Again, the test results ($\chi^2(9) = 9.33$, \emph{p}-value $> 0.406$) did not reject this null hypothesis because $\chi^2$ value with 9 degree of freedom is lower than the critical value at the 0.05 level ($16.92$). This means that there is no association between the level of importance assigned by practitioners to code review practices and the number of systems they developed. To strengthen this result we calculated the effect size and obtained Cramér’s V = 0.008, which also indicates a very weak association. 

Thereby, the analysis of these results allows us to answer RQ2 as follows: 

\medskip
\fbox{\begin{minipage}{30em}
\textit{
Practitioners recognize code review practices as important to the software development process. However, their perception is that companies give less importance to code review practices than they do. This perception may discourage the regular use of these practices.}
\end{minipage}} \linebreak

\subsection{RQ3: What difficulties do developers in Brazil face to use automated static analysis tools to support code review?}

A lot of research effort has been put into improving automated static analysis tools (ASATs) aiming to make code review more objective and standardized. However, a study conducted with 168,241 OSS projects showed that, despite 60\% of the projects make use of ASATs, they typically only use one ASAT in an ad-hoc fashion and not integrated with the flow of development. In addition, the configurations of the ASATs used in those projects barely deviate from the default or introduce custom checks \cite{Beller:2016:SANER}. Our third research question aimed to evaluate the issues and challenges Brazilian practitooners face to use automated static analysis tools.

\begin{figure}
	\centering
	\includegraphics[width=1.0\linewidth]{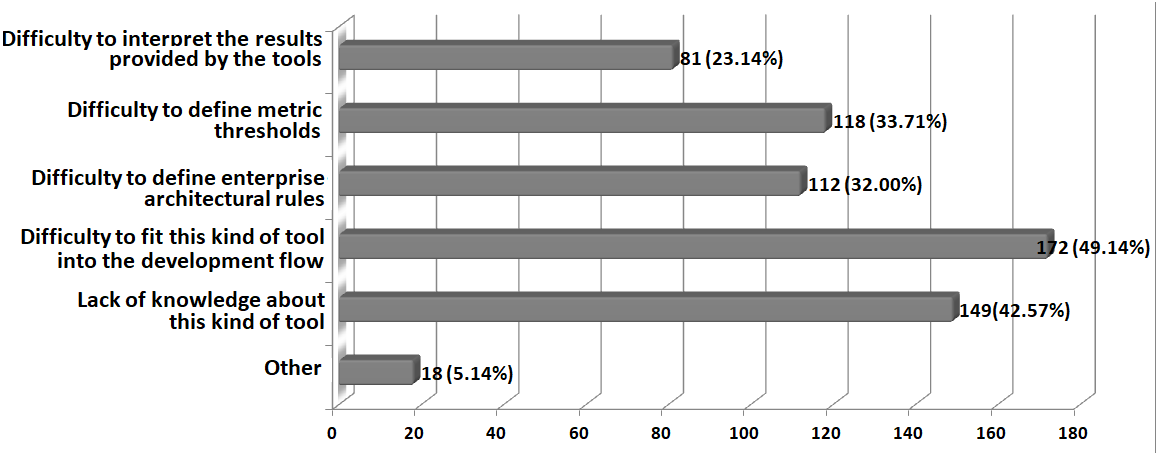}
	\caption{Difficulties to use automated code review tools.}
	\label{fig:codeReviewDifficulties}
\end{figure}

To answer this research question, we only use the fifth question of our survey, which is \emph{What difficulties do you have to use code review tools?} It is a multiple choice question, so that respondents could select more than one difficulty that they consider to hinder their regular use of ASATs. Figure \ref{fig:codeReviewDifficulties} illustrates the obtained results. We discussed in RQ1 that 54.85\% of the respondents claim to use at least one tool to perform code review. We found in RQ3 that 149 out of 350 (42.57\%) respondents declared lack of knowledge about these tools. These results mean that almost everyone who knows ASATs tools uses them in some way. Interestingly, respondents who are unaware of ASATs are scattered across all levels of experience. For instance, regarding the respondents who claimed to have less than two years of experience, 28 out of 112 (25\%) are unaware of ASATs. Other levels of experience have similar rates. For example, among the 216 respondents with two to five years of experience, 50 (23.14\%) are unaware of ASATs. These numbers demonstrate that ASATs need to be better disseminated among practitioners. 

Still regarding difficulties to use ASATs, also a high number of respondents, 172 (49.14\%), claimed to have difficulties to fit the tool into the development flow. Difficulties to fit ASATs into the development flow are also reported by Microsoft developers that use ASATs more regularly \cite{Christakis:2016:ASE}. Developers point out insufficient training and problems to manage large reviews as challenges to fit these tools in development flow \cite{MacLeod:2018:IEEE}. These results show that improving ASATs may not be enough. To reach the benefits promised by the use of ASATs, companies and researchers also need to invest in education and guidelines to fit ASATs into their software development flow. In RQ5 we deep the discussion about the best time of the development flow to apply ASATs.

In addition, a considerable number of respondents, 112 (32\%), claimed to have difficulties to define enterprise architectural rules of the system under review (e.g. rules of communication between architectural layers). Defining the key architectural rules is an essential task for setting parameters, rules and metric thresholds used by many ASATs to perform code review process. This is a challenge also cited by other studies conducted with Microsoft and Mozilla core developers \cite{Kononenko:2016:ICSE, MacLeod:2018:IEEE}. These discussion highlight the need for studies and tools to help developers quickly understand the key architectural decisions of the system under review, such as, architectural layers and their rules of communication. 

A similar number of respondents, 118 (33.71\%), declared difficulties to define metric thresholds (e.g. maximum number of lines of code per method). A number of ASATs allow users to adjust metric thresholds used to identify code anomalies. The accuracy of metric-based assessment is heavily influenced by the calibration of metric thresholds \cite{Sharma:2018:JSS}. Threshold selection is a challenge because of the proneness to false positives \cite{Kessentini:2014:TSE}. A threshold that points out code smells that hold good in the context of an application module may not necessarily make sense for other applications or other modules of the same application \cite{Fontana:2015:WETSOM}. Previous works suggest that deriving metric thresholds according to the application design context might reduce false code smell alarms \cite{Zhang:2013:ICSM,Aniche:2016:SCAM,Dosea:2018:ICPC}. In RQ4 we discussed the developers' perception about considering the context to derive metric thresholds.

Finally, 81 (23.14\%) respondents declared problems to interpret the results pointed out by the tools. This rate seems to be low but we need take into account a high rate of respondents that declared lack of knowledge about ASATs. In fact, the way ASATs present results is considered as one of the main barriers to the consistent and widespread use of ASATs by many studies \cite{Johnson:2013:ICSE, Bacchelli:2013:ICSE, Kononenko:ICSM:2015, Christakis:2016:ASE}. The high number of false alarms, which is also considered as one of the main barriers to ASATs use \cite{Johnson:2013:ICSE, Bacchelli:2013:ICSE, Kononenko:ICSM:2015, Christakis:2016:ASE}, may hinder ASATs results interpretation too. Eighteen respondents (5.14\%) cited other difficulties to use ASATs, including: (i) lack of culture and knowledge of developers and (ii) difficulty use ASATs in legacy code, because it usually comprises design rules different from newer applications. These results also show that training is needed and that ASATs should be adapted to particular contexts. 

Therefore, the analysis of these results allows us to answer RQ3 as follows: 

\medskip
\fbox{\begin{minipage}{30em}
\textit{
Many Brazilian practitioners are still unaware of ASATs or have difficulty to adapt them to their development flow or have difficulty to interpret their results. Thus, research involving ASATs should not only be limited to improving accuracy but must also be concerned with providing guidelines for developers to use them and to adjust them to their particular software development processes.}
\end{minipage}} \linebreak

\subsection{RQ4: What is the developers' perception about evaluating source code using multiple threshold values for each metric?}

A major reason for the occurrence of false positive and negatives on metric-based code smells detection is the lack of context for metric thresholds \cite{Sharma:2018:JSS}. Nevertheless, popular ASATs use generic metric thresholds for the metrics used for detecting code smells. We have a generic threshold for a given metric when we use the same single value for classifying into categories (such as low or high) every class (or every method) of one or more systems. For instance, Lanza and Marinescu \cite{Lanza:2006:oo-metrics} classify as long any method that has more than 20 lines of code (LOC) in Java systems. In this case, 20 is used as a generic threshold for LOC. Using a generic metric threshold for each metric to evaluate all system classes ends up disregarding contextual information of each evaluated class. Some studies have shown that generic metric thresholds might not make sense for the entire set of classes in a system \cite{Lavazza:2016:PROMISE} and taking into account context factors to define multiple context-sensitive thresholds could improve accuracy and reducing false-positive alarms \cite{Zhang:2013:ICSM,Aniche:2016:SCAM, Dosea:2018:ICPC}. RQ4 aims to evaluate the practitioners' perception about context-sensitive metric thresholds to source code evaluation.

\begin{figure}
	\centering
	\includegraphics[width=1.0\linewidth]{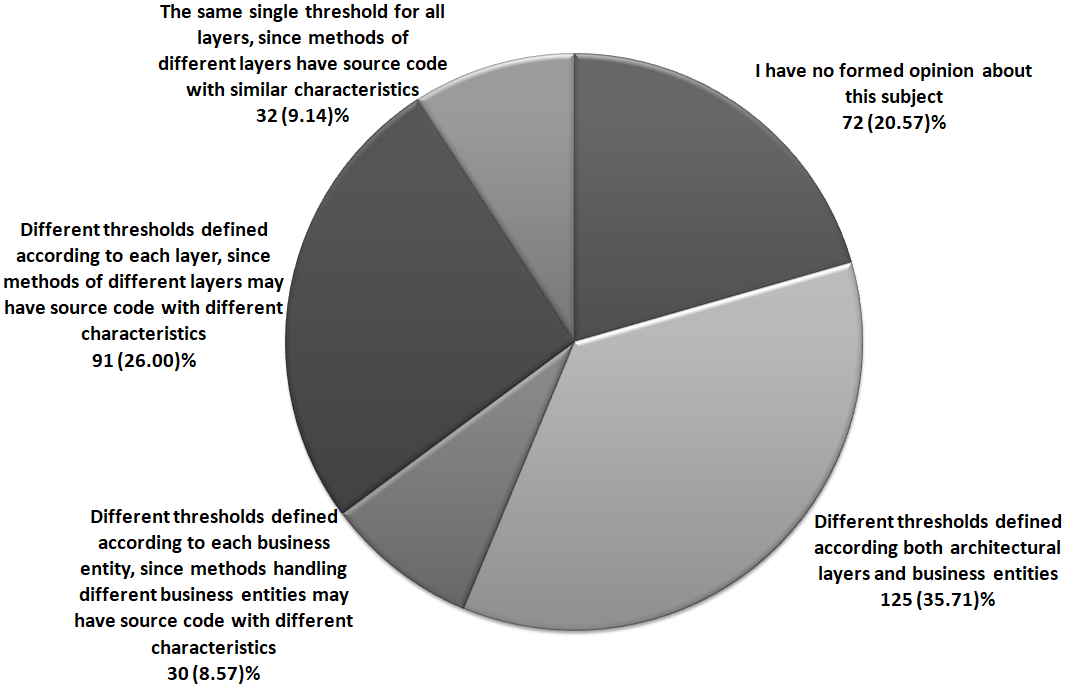}
	\caption{Practitioners' perception of the influence of class context in the selected metric thresholds.}
	\label{fig:codeContext}
\end{figure}

To answer RQ4, we rely on the the sixth question of our survey, which is: \emph{Automated Static Analysis tools usually use a single metric threshold to evaluate all system classes. Considering a three-tier system (GUI, Business and Persistence), what is your opinion about the threshold values that should be used to evaluate classes in each of these three tiers?} It is a single question, so that the respondent is only allowed to choose one answer. The question asks the respondent to consider, as example, a system developed according a three-tier architecture (GUI, Business and Persistence). Then, it asks the opinion of the respondents about whether a single metric thresholds should be used to evaluate the entire source code or different thresholds should be used for different architectural layers or business entities. By business entities we mean the main entities handled by the system. For instance, in a library management system, some examples of business entities are book, book author, publisher, book borrowing, and book return. We hypothesize that some business entities may be more complex (e.g. book borrowing) to be handled by the system and therefore should be evaluated with threshold values different from the ones used for simpler business entities (e.g. book author).

Figure \ref{fig:codeContext} shows that only 32 (9.14\%) respondents declared given a metric, the same single threshold should be used to analyze source code in all three layers. This result shows that practitioners' perception is different from the strategy adopted by most ASATs, which allow only a single threshold for each supported metric. On the other hand, 91 (26.0\%) respondents claimed that metric thresholds should be different and defined according each architectural layer, since the source code in methods of one layer may have distinct characteristics from methods in the other layers. Also, 30 (8.57\%) respondents declared that metric thresholds must be different and defined according to each business entity, since methods that handle different business entities may have different characteristics. Moreover, 125 (35.71\%) respondents believe that different metric thresholds should be defined according both architectural layer and business entities. In summary, 246 out of 350 (70.28\%) respondents do not agree with the use of single generic thresholds. Finally, 72 (20.57\%) respondents did not have a formed opinion regarding this subject. These results motivate future research for investigating whether using multiple metric thresholds reduces the number of false alarms current ASATs return.

We also investigated if these results are associated to practitioners' work experience or the number of system they developed. First, we cross-analyzed the answers applying the chi-square independence test aiming to verify whether the following null hypothesis is rejected: \emph{There is no association between practitioners' perception about multiple thresholds and practitioners' work experience}. The test results were $\chi^2(12) = 15.14$ and \emph{p}-value $>0.23$, which did not reject the null hypothesis because $\chi^2$ value with 12 degree of freedom is lower than the critical value at the 0.05 level ($21.02$). This means that there is no association between practitioners' perception about multiple thresholds and practitioners' work experience. To strengthen this result we calculated the effect size and obtained Cramér’s V = 0.014, which indicates a very weak association. Second, we cross-analyzed the answers applying the chi-square independence test aiming to verify whether the following null hypothesis is rejected: \emph{There is no association between practitioners' perception about multiple thresholds and number systems the practitioners developed}. Again, the test result did not reject the null hypothesis, showing no association between the two variables ($\chi^2(12) = 13.80$, \emph{p}-value $>0.31$) because $\chi^2$ value with 12 degree of freedom is lower than the critical value at the 0.05 level ($21.02$). To strengthen this result we calculated effect size and obtained Cramér’s V = 0.013, which indicates also a very weak association. 

In summary, we observed only few respondents that agree with single generic metric thresholds usually adopted by the most popular ASATs. This allow us to answer RQ4 as follows: 

\medskip
\fbox{\begin{minipage}{30em}
\textit{
Instead of a single generic metric threshold, Brazilian practitioners believe that ASATs should use multiple thresholds, calibrated according contextual design information, such as architectural layers. Thus, future research should further investigate the use of multiple thresholds.}
\end{minipage}} \linebreak

\subsection{RQ5: What is the best time in the software development process to conduct automatic code review practices?}

A challenge reported by many developers is selecting the best time to apply code review practices in the software development process \cite{Ayewah:2008:JIE, Johnson:2013:ICSE, Christakis:2016:ASE}. ASATs usually allow code analysis directly by developers \cite{Johnson:2013:ICSE}. In addition, some tools allow the analysis to be launched by an integration server \cite{Sadowski:2015:ICSE}. Recent studies propose to show immediate feedback while developers program \cite{Do:2017:ISSTA}, but this feature is unavailable in most of the existing ASATs. In RQ5 we aim to obtain a large-scale perception of practitioners about the best time to automatically check code anomalies. This result is important to guide future research and the development of tools that allow code analysis to be performed in multiples moments.

\begin{figure}
	\centering
	\includegraphics[width=1.0\linewidth]{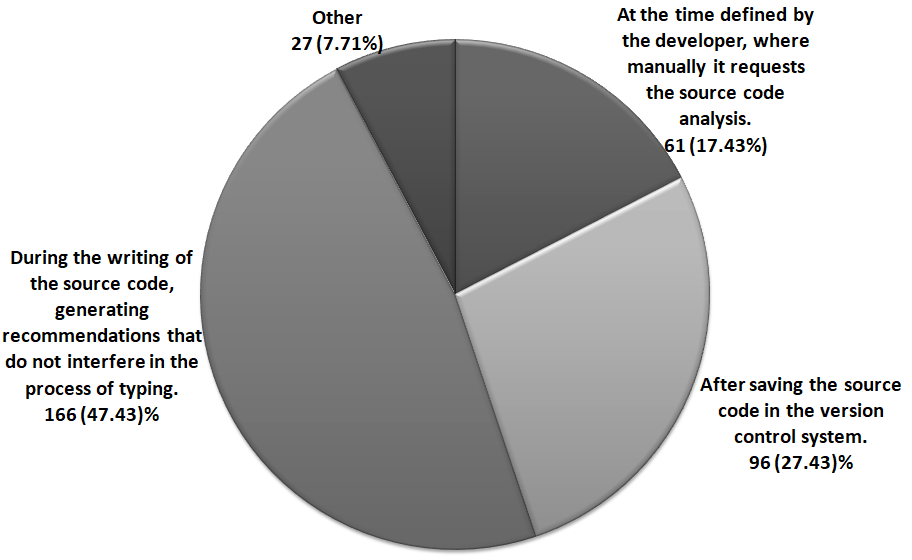}
	\caption{The best time to apply automatic code review practices.}
	\label{fig:codeReviewtime}
\end{figure}

To answer RQ5, we had in our survey the following question, which allows single response: \emph{What is the best moment to warn software developers regarding code anomalies?}. Figure \ref{fig:codeReviewtime} summarizes the answers to the survey question. It shows that 166 out of 350 (47.43\%) respondents claimed that automatic code review should be applied while developers write source code, generating, however, recommendations that do not hinder on the typing process. On the other hand, 96 (27.43\%) respondents declared that automatic code review should be applied after saving the source code in the version control system. Also, 61 (17.43\%) respondents claimed that automatic code review should be applied at the time demanded by the developer, who manually requests the tool to proceed with the source code analysis. Finally, 27 (7.71\%) respondents suggested other moments to conduct automatic code review, for instance, during test phase. 
We also investigated if these results are associated to practitioners' work experience and the number of systems they developed. First, we cross-analyzed the answers applying the chi-square independence test aiming to verify whether the following null hypothesis is rejected: \emph{There is no association between practitioners' perception about the best time to apply ASATs and practitioners' work experience}. The test results ($\chi^2(9) = 10.77$, \emph{p}-value $>0.29$) did not reject this null hypothesis because $\chi^2$ value with 9 degree of freedom is lower than the critical value at the 0.05 level ($16.92$). This means that there is no association between practitioners' perception about the best time to apply ASATs and practitioners' work experience. To strengthen this result we calculated the effect size and obtained the Cramér’s V = 0.010, which indicates a very weak association. Second, we cross-analyzed the answers applying the chi-square independence test aiming to verify whether the following null hypothesis is rejected: \emph{There is no association between practitioners' perception about the best time to apply ASATs and the number of systems they developed}. The test results were $\chi^2(9) = 18.35$ and \emph{p}-value $=0.031$. In this case, the results rejected the null hypothesis because $\chi^2$ value with 9 degree of freedom is larger than the critical value at the 0.05 level ($16.92$). This means that there is an association between practitioners' perception about the best time to apply ASATs and the number of systems they developed. However, Cramér’s V is 0.017, which indicates a very weak association. 

In most popular existing ASATs, developers are responsible for triggering source code analysis whenever they want to do so. However, the survey results show that 47.43\% of the respondents would like to have immediate feedback, which allow them to perform repair while they write source code. It seems that if ASATs only warn them about any problem they introduce in the source code late, they will be less willing to return to the problematic point to repair it. Previous studies also showed that developers usually are concerned with different categories of defects while carrying out different tasks, such as writing code, reviewing code or integrating code to an integration server \cite{Vassallo:2018:SANER}. Future researches should be conducted to investigate whether automatic source code analysis is really useful during these three moments and what categories of warnings ASATs should detect during each task.

In summary, we observed that respondents have distinct time preferences to perform automatic code review. This allows us to answer RQ5 as follows: 

\medskip
\fbox{\begin{minipage}{30em}
\textit{
Most practitioners claimed that the best time to perform automatic code reviews is while they write code. However, other practitioners mentioned that ASATs should allow developers to execute them whenever they request it or should allow analysis just after source code is saved on the version control system. Therefore, ASATs should allow source code analysis in different moments. This could help developers to fit such tools into their specific development processes. Also, this would allow future researches about what categories of warning best fit into different software development phases.}
\end{minipage}} \linebreak
\section{Threats to Validity}\label{sec:threats}

In general on-line surveys are considered to have lower internal validity and stronger external validity in comparison with other means of empirical investigation, such as case-studies or experiments \cite{Punter:2003:ISESE}. We discuss the threats to the validity of our study according to four categories \cite{Wohlin:2012:book}:

\emph{Construct validity.} In a survey, such threats may
mainly occur because respondents could possibly interpret a question in a different way than it has been conceived, possibly producing misleading results. To minimize this threat, as explained in Section \ref{sec:studySettings}, we tested the questionnaire, by means of two pilot studies, to check possible problems related to ambiguity, missing response options, and lack of clarity in our questions. Also, to make important concepts and terms (e.g., static analysis tools) clear, we included examples of them in the questions or answer options.

\emph{Internal validity.} Threats to internal validity are related to issues that may affect the causal relationship between treatment and outcome. In general, it is hard to control these factors since survey is an unsupervised study and the level of control is very low. To avoid apprehension, we guaranteed the respondents their complete anonymity. Also, there is always a risk that different respondent backgrounds (e.g., experience) influence the experiment results. However, due to the large sample, as well as the range of competence and experience levels, this risk was limited. Similarly, the large sample mitigated any threat potentially caused by respondents with different personalities \cite{Feldt:2010:IST}.

\emph{External validity.} Despite the size of our sample being large enough to enable statistically significant results, we are also aware that the sample size could not be large enough for generalization purposes given the lack of accurate data about the target population. However, our sample is similar to other surveys conducted on different software engineering subjects \cite{Christakis:2016:ASE, Bosu:2017:TSE}. In addition, we do not claim that our conclusions can be generalized outside the scope of our study.

\emph{Conclusion validity.} Threats to conclusion validity are concerned whether correct conclusions are reached through rigorous and repeatable treatment \cite{Wohlin:2012:book}. To minimize possible errors related the target audience sampling, we used, for each research question, non-parametric statistical tests and measured the effect size to discuss our findings. Therefore, all the conclusions that we drew in this survey are strictly traceable to data. Moreover, to increase transparency, the survey data is available online\footnote{http://dx.doi.org/10.17632/gzz6wmp66j.3} so that other researchers can validate it or replicate the study.
\section{Related Works}\label{sec:relatedWorks}

Some studies capture the perception of practitioners who already use code review practices regularly. Differently, we conducted a large scale survey without limiting the target audience to code review experts. With this, we expected to capture others issues and challenges about code review adoption. In addition, we also attempted to capture the practitioners' perception about multiple metric thresholds that take the context of source code elements into account. Using contextual metric thresholds is a possible way to avoid what many software developers consider as a key problem on the use of ASATs: the high number of false alarms. Previous studies do not address this point.

An initial small-scale study conducted by Johnson \emph{et al.} \cite{Johnson:2013:ICSE} investigated 20 developers using semi-structured interviews to know why static analysis tools are not widely used and how these tools could be improved to increase usage based on developer feedback. The study focused on static analysis tools that have well-defined programming rules to find defects (e.g. FindBugs, Lint, IntelliJ, and PMD). Most of the developers (19 of 20) claimed that static analysis tools do not present their results with enough information that allows them to clearly understand the problem and know what they should be doing differently. The same number of developers expressed the importance of offering different ways to fit a tool into the software development process. Some developers prefer finding a “stopping point” in their code to run the tool \cite{Layman:2007:ESEM}. Other developers prefer the tool running in the background. Our study also investigated how developers prefer to use code review tools, but based on the opinion of a considerably higher number of developers. In addition, we also investigated other issues that may discourage developers to adopt static analysis tools.

Bacchelli and Bird \cite{Bacchelli:2013:ICSE} conducted an exploratory study following a mixed approach and collecting data from different sources for triangulation. They (i) observed 17 industrial developers performing code review; (ii) interviewed these developers using a semi-structured interview; (iii) manually inspected and classified the content of 570 comments in discussions about code reviews; and (iv) surveyed 165 managers and 873 programmers. The results showed that finding defects and code improvement are the primary motivations to code review, although participants believe that code review brings other benefits, for example, knowledge transfer and proposition of alternative solutions. They also identified when the business context of the software is clear and understanding is very high, as in the case when the reviewer is the owner of changed files, code review comments have better quality. Bosu et al. \cite{Bosu:2017:TSE} conducted a survey with 416 Microsoft developers aiming to provide additional insight into similarities or differences between OSS and Microsoft developers. They aim to verify if Microsoft developers who work on distributed projects would have similar views about code as OSS developers (whose projects are also distributed). The results show a large amount of similarity between the Microsoft and OSS respondents and a little difference between distributed and co-located Microsoft teams. They also verify that developers spend approximately 10-15 percent of their time in code reviews, with the amount of effort increasing with experience. Our survey complements these two studies as it identifies additional issues and challenges developers face for adopting code review practices, such as problems to fit code review tools in their software development process, difficulties to define metric thresholds and difficulties to interpret ASAT results.

Beller at al. \cite{Beller:2016:SANER} conducted a study to understand the prevalence of ASATs, their configuration in real software projects, and how those configurations evolve over time. Firstly they analyzed the use of nine popular ASATs in 122 Open-Source Software (OSS) projects. Then, they analyzed how ASATs were configured and how their configuration settings evolved in 168,241 OSS projects. The results show that 60\% of the most popular and (therefore arguably) most advanced projects make use of ASATs, although they typically use only one ASAT in an ad-hoc fashion, not integrated with the flow of development. Regarding to ASAT configurations, the study showed that, after an one-week period of changes, the ASAT configurations usually remain unchanged along the rest of the project. Additionally, ASAT configurations barely deviate from the default settings and rarely comprise custom checks. Our study evaluated developers' perception about new ideas related to ASAT configurations, such as, the use of multiple metric thresholds.

Kononenko et al. \cite{Kononenko:2016:ICSE} performed a survey with 88 Mozilla core developers to understand their challenges regarding code review. They reported key technical challenges, such as gaining familiarity with the code, coping with code complexity, and having suitable tool support. And they also reported personal challenges related to time management, technical skills, and context switching. We also investigated challenges related to code review and raised additional points, such as lack of knowledge about static analysis tools and difficulties to fit them into the software development process.

MacLeod et al. \cite{MacLeod:2018:IEEE} conducted semi-structured interviews with 18 developers from four teams at Microsoft. The initial findings about tool use, developer motivations, and the challenges developers face were validated through a survey with 911 Microsoft developers. The study revealed that Microsoft developers recognize code reviews’ value and importance. They appreciate reviewer feedback and they develop more thorough artifacts when they know that someone will revise them. The study showed that 87\% of the respondents acted as a code reviewer during the previous week of the research. Improve the code, find defects, transfer knowledge and explore alternative solutions are ranked as the main motivations for code reviews. Reviewers point out the main challenge to conducted code reviews are the difficulty to manage large reviews, finding time to do reviews, understanding a changing and its motivation, finding relevant documentation and understanding the history of changes and decisions. They also complained about insufficient training for reviews. Christakis and Bird \cite{Christakis:2016:ASE} interviewed and surveyed developers across Microsoft to understand their needs and how ASATs can or do fit into their process. They also examined many corrected defects to understand what types of issues occur most and least often. They received 375 responses to the survey, yielding a 19\% response rate. Developers point out the main pain points, obstacles, and challenges to use ASATs are (i) wrong checks are on by default, (ii) bad warning messages, (iii) too many false positives, (iv) too slow, (v) no suggested fixes, (iv) difficult to fit into the workflow. Developers also prefer that ASATs show alerts in the code editor followed by the build output. In addition, developers suggested that ASATs should have a false positive rate no higher that 15\% to 20\%. These results are in line with the findings of previous works \cite{Johnson:2013:ICSE, Ayewah:2008:JIE}. Our target audience also recognizes code review as important, but only very few respondents claimed to perform code review regularly. Also, as recent studies showed that context-sensitive metric thresholds could decrease the number of false positives \cite{Zhang:2013:ICSM, Aniche:2016:SCAM}, our study goes further on this point by investigating developers' perception about multiple metric thresholds.

Vassalo et al. \cite{Vassallo:2018:SANER} argue that developers tend to fix different warnings in different stages and, therefore, the development activity being performed or the context of ASAT usage are factors that should be considered to improve warning prioritization. They first explored the adoption of ASATs in practice through a survey with 42 developers (69\% working in the industry and 31\% open source contributors) that integrate ASATs in their software release pipeline. Then, semi-structural interviews with 11 industrial developers reinforced their initial findings. The results showed that 37\% of respondents rely on ASATs while integrating code changes in an existing project, 33\% while reviewing code and 30\% while working locally. Therefore, ASATs are adopted in three main development contexts: (i) local environment, (ii) code review, and (iii) continuous integration. They also discovered that 51\% of the respondents configure ASATs at least once before starting a new project. Also, 75\% of the respondents declared to not make distinct configurations to use ASATs in these three different contexts. Finally, they observed that developers usually pay attention to different categories of defects while working locally, during code review or rather in continuous integration. However, current ASATs are not able to show different categories of warnings according to different development contexts. Our study also shows that most Brazilian practitioners would like to evaluate code problems while writing source code. But others prefer other phases of the development process. ASATs need, therefore, to be designed to fit different stages of the development process.

In summary, as our survey target audience was not limited to code review experts, we were able to identify some challenges different from those related work identified. These additional challenges give new insights for future studies, such as studies to evaluate the best moments of the software development process to apply code review and studies to investigate metric thresholds that take source code context into account.
\section{Conclusion}\label{sec:conclusion}

We conducted a web-based survey with 350 Brazilians practitioners engaged in the software industry whose code review practices are not so well established. The survey featured questions about (i) the practices used to code review, (ii) the importance given to such practices, (iii) the issues about the automation of these practices (iv) practitioners' perception about multiples metric thresholds, and (v) the best time to perform automatic code review. The results and their implications for research and practice can be summarized as follow.

\emph{Code review practices are known but applied irregularly.} Our results showed that Brazilian practitioners know code review practices (RQ1) and recognize their importance (RQ2). However, development teams do not apply these practices on a regular basis (RQ1). A potential implication of this is that future researches should conduct studies to clarify the impact and benefits of applying code review practices in different phases of the software development process. Additionally, researchers should propose guidelines for industry practitioners to use these practices.

\emph{Practitioners are unaware or have many issues to use automated static analysis tools.} Our results showed that 54.85\% of respondents use at least one automatic code review practices (RQ1). However, many respondents (42.85\%) stated unaware of these tools (RQ3). Additionally, practitioners reported many issues and challenges to adopt these tools regularly (RQ3). 
A potential implication of these results is that future researches should study and propose solutions for integrating such tools in software development processes more easily.

\emph{Practitioners' perception is that multiple metric thresholds should be used in source code quality analysis.} Current static analysis tools use a single metric threshold value for each metric. Metric thresholds are used by detection strategies to identify code anomalies. However, our results showed that only 9.14\% of the respondents declared to agree with a single metric threshold to evaluate all system classes (RQ4). On the other hand, 70.28\% of them believe that contextual factors, such as architectural layers or business entities, influence metric values, so that multiple thresholds should be defined for each metric (RQ4). A potential implication of this is that future researches should propose techniques for defining multiple context-sensitive thresholds and investigate whether they improve static analysis tool accuracy.

\emph{Static analysis tools must be designed to run at different stages of the software development process.} 
Current static analysis tools usually perform source code analysis when they are manually triggered by the developer. Some tools also allow performing analysis in the integration server. However, our results showed that most software developers (47.43\%) would like to perform source code analysis while they write source code. An implication of these results is tool developers should build static analysis tools that are flexible and fulfill different requirements of code analysis depending on the stage of development process they are applied. 

Finally, as lessons learned, our study illustrates that surveys with a well-defined focus, language close to the respondents and that can be answered quickly have a high response rate. Also, the use of examples to clarify terms and concepts that are not always clear to the respondents proved to be an important tool for the developers' comprehension of survey questions.

\section*{Acknowledgements}
This work was supported by CNPq (grant 312153/2016-3).





\bibliographystyle{model1-num-names}
\bibliography{biblio}







\end{document}